\documentclass[
    ,final            
  ]
  {aipproc}

\layoutstyle{6x9}

\begin{document}

\title{Generalised thermostatistics using hyperensembles}


\author{Jan Naudts}{
  address={Department of Physics, University of Antwerp, Groenenborgerlaan 171, 2020 Antwerpen, Belgium}
}

\begin{abstract}
The hyperensembles, introduced by Crooks in a context of non-equilibrium statistical
physics, are considered here as a tool for systems in equilibrium.
Simple examples like the ideal gas, the mean-field model, and the
Ising interaction on small square lattices, are worked out to illustrate the concepts.
\end{abstract}

\maketitle


\newcommand{\be}{\begin{eqnarray}}
\newcommand{\ee}{\end{eqnarray}}


\section {Introduction}

The starting point of \emph {superstatistics} \cite {BC03,BC07} is the assumption that
the inverse temperature $\beta$ of the Boltzmann-Gibbs distribution
\be
q_\beta(x)=\frac 1{Z(\beta)}e^{-\beta H(x)}
\ee
may itself be a fluctuating quantity, governed by a probability density $f(\beta)$.
The resulting probability distribution is then determined by
\be
p_f(x)=\int_0^\infty{\rm d}\beta\,f(\beta)q_\beta(x).
\ee
The choice of the \emph {hyperdistribution} $f(\beta)$
(called the \emph {entropic distribution} in \cite {CGE07}) depends on the application at hand.
Recently, Crooks \cite {CGE07} suggested that this distribution
should be determined by means of the Maximum Entropy Principle (MEP)
because, in general, information is lacking about what choice of $f(\beta)$ is appropriate.
Additional constraints, introduced when applying the MEP, then lead to a
parametrised family of probability distributions. The latter he called a
\emph {hyperensemble}.

The choice of constraints is crucial. Crooks proposes to use three
constraints: the normalisation, the mean energy and the mean entropy.
The result of applying the MEP is that the hyperdistribution is of the form
\be
f(\beta)=\frac {c(\beta)}{Z(\theta,\lambda)}
\exp\left(\lambda\check S(\beta)-\theta\langle H\rangle_\beta\right).
\ee
Here, $\theta$ is the Lagrange multiplier controlling the mean energy,
$\lambda$ controls the mean entropy.
The canonical entropy is denoted $\check S(\beta)$ to distinguish it from the
entropy of the hyperensemble.
The normalisation is given by
\be
Z(\theta,\lambda)=\int_0^\infty c(\beta){\rm d}\beta\,\exp\left(\lambda\check S(\beta)-\theta\langle H\rangle_\beta\right).
\ee

The intention of Crooks is to use the notion of hyperensembles to study systems out of
equilibrium. The present paper shows the usefulness of hyperensembles as a unifying
concept for systems in equilibrium. In this context the parameter $\lambda$ plays no important
role and may be taken equal to 1.


\section {The ideal gas}

Consider only the momenta $p_1,p_2,\cdots,p_N$ of an ideal gas containing $N$ particles.
The Hamiltonian is
\be
H(p)=\frac 1{2m}\sum_{n=1}^N\sum_{\alpha=1}^3p_{n\alpha}^2.
\ee
The canonical probability distribution is
\be
q_\beta(p)=\frac 1{(2\pi mk_BT)^{3N/2}}e^{-\beta H(p)}.
\ee
The entropy and energy equal
\be
\check S(\beta)=-\frac {3N}2\ln\beta,
\qquad\qquad
\langle H\rangle_\beta=\frac {3N}2\beta^{-1}.
\ee
After optimisation, the hyperdistribution becomes
\be
f(\beta)=\frac {c(\beta)}{Z(\theta)}\exp\left(-\frac {3N}2\ln\beta-\theta\frac {3N}{2\beta}\right).
\ee
This is the \emph {inverse Gamma distribution}, also called the inverse $\chi^2$-distribution. It has a maximum at $\beta=\theta$.
Hence, the canonical probability distribution $q_\beta(p)$, with $\beta=\theta$,
is the most likely distribution at inverse hypertemperature $\theta$.

The ideal gas, together with the above hyperdistribution, was already considered in the context
of superstatistics \cite {TH04,TB05}.

\section {The mean-field approximation}

\begin{figure}
\label {fig:mf}
\includegraphics[width=6cm] {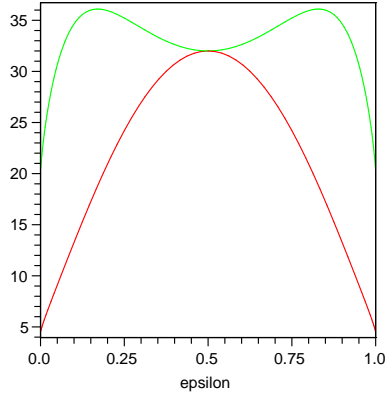}
\caption{
Hyperdistribution $f_{\beta,h}(\epsilon)$ of the mean-field model with $J=1$,
$N=5$, $h=0$, and $\beta=0.6$ and $\beta=1.2$, respectively.
}
\end{figure}

Consider spin variables $\sigma_1,\sigma_2,\cdots,\sigma_n$, taking on values $\pm 1$.
The equilibrium states of the ideal paramagnet are the product states
\be
q_\epsilon(\sigma)=\prod_{n=1}^N\epsilon^{\frac 12(1+\sigma_n)}(1-\epsilon)^{\frac 12(1-\sigma_n)},
\ee
with $0\le \epsilon\le 1$.
The total magnetisation is $\displaystyle \langle M\rangle_\epsilon = N(2\epsilon -1 )$.
The entropy is
\be
\check S(\epsilon)=-N\epsilon\ln\epsilon - N(1-\epsilon)\ln (1-\epsilon).
\ee
Here, we use $\epsilon$ as a parameter instead of $\beta$.
Introduce the ferromagnetic Hamiltonian
\be
H=-\frac 12\sum_{m,n=1}^NJ_{m,n}\sigma_m\sigma_n-h\sum_{m=1}^N\sigma_m.
\ee
Its average, using the equilibrium distribution of the paramagnet and assuming $J_{mm}=0$,
equals $\langle H\rangle_\epsilon=-\frac 12NJ(2\epsilon-1)^2-hN(2\epsilon-1)$, with
\be
J=\frac 1{N}\sum_{m,n=1}^NJ_{m,n}.
\ee
The optimised hyperdistribution with parameters $\beta$ (instead of $\theta$) and $h$ is
\be
& &f_{\beta,h}(\epsilon)
=\frac 1{Z(\beta,h)}
\exp\left( \check S(\epsilon)-\beta \langle H\rangle_\epsilon\right)\crcr
&=&\frac 1{Z(\beta,h)}\exp\bigg( N\bigg[
-\epsilon\ln\epsilon - (1-\epsilon)\ln (1-\epsilon)
+\frac 12\beta J(2\epsilon-1)^2+\beta h(2\epsilon-1)\bigg]\bigg).\crcr
& &
\ee
For $\beta>\beta_{\rm mf}=J^{-1}$ and $h=0$ the hyperdistribution has two maxima,
corresponding to two equally likely states. See the Figure \ref {fig:mf}.
At high temperature the paramagnetic state $q_\epsilon(\sigma)$ with $\epsilon=1/2$
is the most likely state. At low temperature the two states with $\epsilon\not=1/2$, maximising $f_{\beta,h}(\epsilon)$,
are the most likely ones.
This bifurcation from a paramagnetic state to a pair of ferromagnetic states
as a function of the temperature $\beta^{-1}$ involves a spontaneous magnetisation of the system.

In the quantum case the ideal paramagnet is described by a density matrix 
of the product form $\rho_\theta^{\otimes N}$, with
\be
\rho_\theta=\frac 12+\sum_{k=1}^3\theta_k\sigma_k.
\ee
The $\sigma_k$ are the Pauli matrices.
The Hamiltonian of the Heisenberg model reads
\be
H=-\frac 12\sum_{m,n=1}^{N}J_{m,n}\sum_{k=1}^3\sigma_{mk}\sigma_{nk}
-h\sum_{n=1}^N\sigma_{n3}.
\ee
The matrix $\sigma_{mk}$ is a copy of $\sigma_k$ at site $m$.
Without restriction, assume $J_{nn}=0$.
Then the average energy in the product state is $\langle H\rangle_\theta=-NJ|\theta|^2+Nh\theta_3$,
with $J$ as in the classical case. The von Neumann entropy of the product state equals
\be
S\left(\rho_\theta^{\otimes N}\right)
&=&-\frac N2(1+|\theta|)\ln \frac 12(1+|\theta|)-\frac N2(1-|\theta|)\ln \frac 12(1-|\theta|).
\ee
One can now calculate the hyperdistribution $f_{\beta,h}(\theta)$. Note that
this is a classical distribution, not an operator. A maximum is attained
when the following three equations are satisfied
\be
\frac 12\frac {\theta_k}{|\theta|}\ln\frac {1+|\theta|}{1-|\theta|}=-2\beta J\theta_k-h\delta_{k,3},
\quad k=1,2,3.
\ee
Assuming $h\not=0$, the solution requires $\theta_2=\theta_3=0$.
The third parameter $\theta_3$ must be a solution of the
usual mean-field equation $\theta_3=\tanh(\beta J\theta_3+\theta_3 h)$.

\section {The microcanonical ensemble}

Following Boltzmann, the probability distribution of the microcanonical
ensemble equals
\be
q_E(x)=\frac {c(x)}{\rho(E)}\delta(E-H(x)),
\ee
with the density of states $\rho(E)$ given by
$ \displaystyle \rho(E)=\int{\rm d}x\,c(x)\delta(E-H(x))$.
The Boltzmann entropy is $\displaystyle \check S(E)=k_B\ln\rho(E)$.
The optimised hyperdistribution becomes
\be
f_\theta(E)
=\frac {1}{Z(\theta)}\exp\left(\check{\cal S}(E)-\theta E\right)
=\frac {\rho(E)}{Z(\theta)}e^{-\theta E}.
\ee
The probability distribution $p_\theta(x)$ of the hyperensemble
coincides with the canonical Boltzmann-Gibbs distribution
\be
p_\theta(x)&\equiv&
\int_0^\infty{\rm d}E\,f_\theta(E)q_E(x)\crcr
&=&\int_0^\infty{\rm d}E\,\frac {\rho(E)}{Z(\theta)}e^{-\theta E}\frac {c(x)}{\rho(E)}\delta(E-H(x))\crcr
&=&\frac {c(x)}{Z(\theta)}e^{-\theta H(x)}.
\ee
The parameter $\theta$ is the inverse temperature $\beta$.

\begin{figure}[!t]
\label {ising:counts}
\caption{
Number of configurations of a 2x2 and a 3x3 Ising lattice as a function of energy and magnetisation.
On the vertical axis is the number of up spins $m$, on the horizontal axis the number of non-matching
neighbour pairs $n$.
}
\includegraphics[width=3cm] {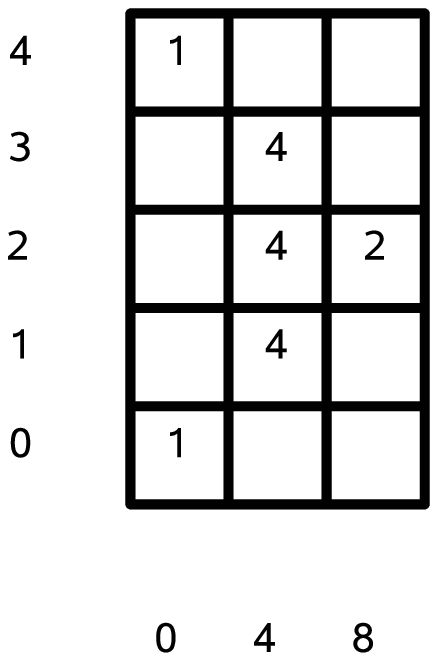}\hskip 2cm
\includegraphics[width=6cm] {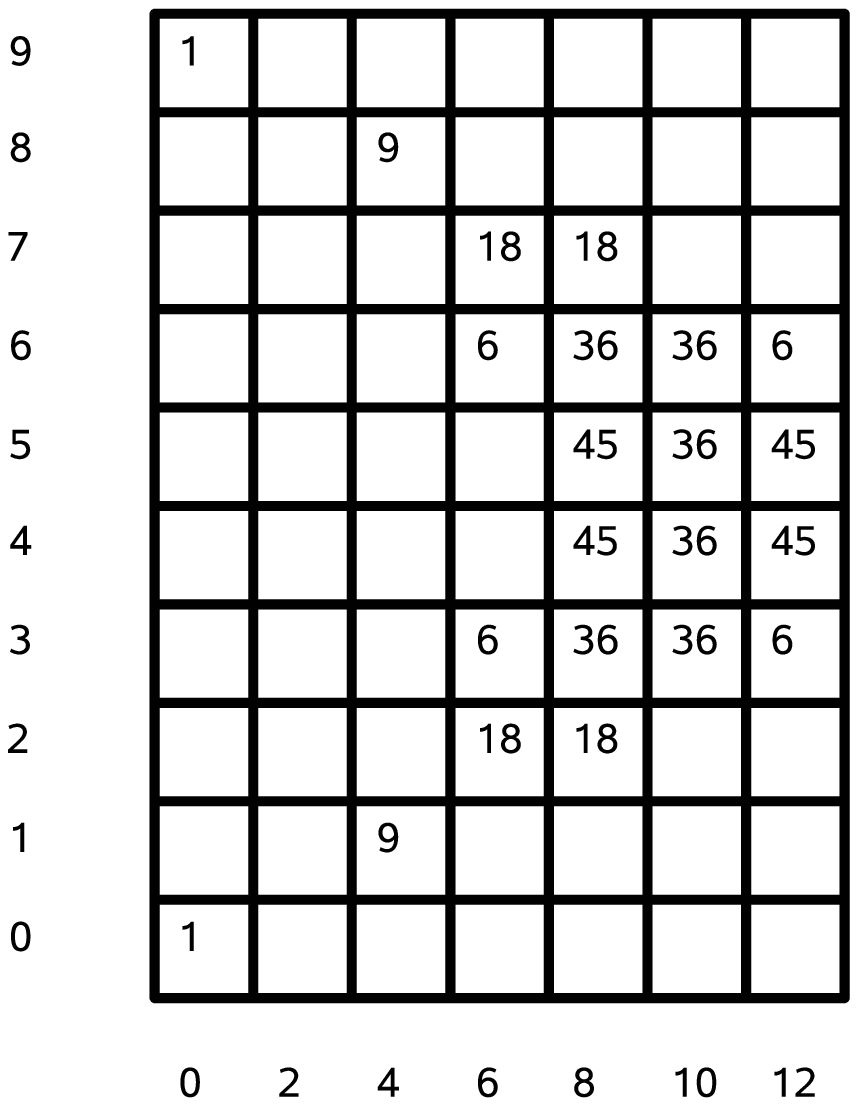}
\end{figure}

\begin{figure}[!ht]
\label {ising:hyperdis}
\includegraphics[width=6cm] {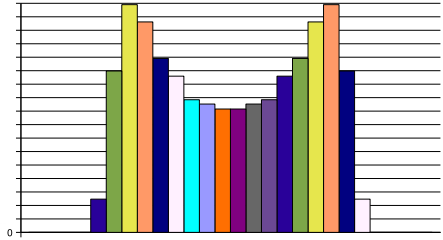}
\caption {
Normalised $C(m,n)$ as a function of $m$ at constant $n=16$ for a 5x5 Ising lattice.
}
\end{figure}

As an example, involving a discrete phase space, let us consider the 2-dimensional Ising model
on a square lattice of $n\times n$ Ising spins. 
Count the number of configurations $C(m,n)$ with $m$ spins up
and $n$ pairs of unequal nearest neighbours. See the Figure \ref {ising:counts}.
The entropy is $\displaystyle \check{\cal S}(m,n)=\ln C(m,n)$.
The hyperdistribution is
\be
f_{\beta,h}(m,n)=\frac {C(m,n)}{Z(\beta,h)}\exp\left(2\beta J(N-n)-\beta h(N-2m)\right).
\ee
At constant $n$ and $h=0$ it is proportional to $C(m,n)$. See the Figure \ref {ising:hyperdis}.
It shows two maxima, corresponding with the spin up and spin down phases.

\section {Discussion}

The aim of the paper is to show that the concept of hyperensembles,
as introduced by Crooks \cite {CGE07} in a context of non-equilibrium systems,
fits well into the standard theory of equilibrium statistical physics.
Several situations have been considered.
In the example of the ideal gas the hyperensemble approach coincides with
the superstatistical treatment of \cite {TH04,TB05}.
In the mean-field theory one starts from a simple model, such as
the ideal paramagnet, to solve more complex models under the constraint that the state
of the system is an equilibrium state of the simple model.
The present formulation in terms of hyperensembles is a mere reformulation of
what is known since long. However, it can be applied in
a very general context. Such an application beyond the traditional
scope of mean-field theory can be found in \cite {VdSN06},
where a random walk model is used to model polymer behaviour.
In the microcanonical ensemble the probability distribution of the hyperensemble 
coincides with the canonical Boltzmann-Gibbs distribution.
The maximum of the hyperdistribution may be degenerate.
This is for instance the case in finite spin lattices with ferromagnetic Ising
interaction. Hence, the most-likely microcanonical state is non-unique.
Such a feature also occurs in mean-filed models and may be interpreted
as a precursor of the phase transition occurring in the thermodynamic limit.




\end{document}